\documentclass[sigconf]{acmart}
\usepackage{cleveref}
\usepackage{array}
\usepackage[shortlabels]{enumitem}
\usepackage{caption}
\usepackage{subcaption}

\AtBeginDocument{%
  \providecommand\BibTeX{{%
    \normalfont B\kern-0.5em{\scshape i\kern-0.25em b}\kern-0.8em\TeX}}}

\copyrightyear{2023}
\acmYear{2023}
\setcopyright{acmlicensed}\acmConference[VRST 2023]{29th ACM Symposium on Virtual Reality Software and Technology}{October 9--11, 2023}{Christchurch, New Zealand}
\acmBooktitle{29th ACM Symposium on Virtual Reality Software and Technology (VRST 2023), October 9--11, 2023, Christchurch, New Zealand}
\acmPrice{15.00}
\acmDOI{10.1145/3611659.3615685}
\acmISBN{979-8-4007-0328-7/23/10}

\acmSubmissionID{1045}

\newcommand{\etal}{\textit{et al}. }
\newcommand{\ie}{\textit{i}.\textit{e}. }
\newcommand{\eg}{\textit{e}.\textit{g}. }

\begin{document}
\title[Evaluating Augmented Reality Communication: How Can We Teach Procedural Skill in AR?]{Evaluating Augmented Reality Communication:\\How Can We Teach Procedural Skill in AR?}


\author{Manuel Rebol}
\email{rebol@gwu.edu}
\affiliation{%
  \institution{Graz University of Technology}
  \city{Graz}
  \country{Austria}
}
\affiliation{%
  \institution{American University}
    \city{Washington}
  \state{DC}
  \country{USA}
}
\affiliation{%
  \institution{George Washington University}
  \city{Washington}
  \state{DC}
  \country{USA}
}

\author{Krzysztof Pietroszek}
\email{pietrosz@american.edu}
\affiliation{%
 \institution{American University}
  \city{Washington}
  \state{DC}
  \country{USA}}


\author{Neal Sikka}
\email{nsikka@mfa.gwu.edu}
\author{Claudia Ranniger}
\email{cranniger@mfa.gwu.edu}
\author{Colton Hood}
\email{chood@mfa.gwu.edu}
\affiliation{%
  \institution{George Washington University}
  \city{Washington}
  \state{DC}
  \country{USA}
}

\author{Adam Rutenberg}
\email{arutenberg@mfa.gwu.edu}
\affiliation{%
  \institution{George Washington University}
  \city{Washington}
  \state{DC}
  \country{USA}
}

\author{Puja Sasankan}
\email{pujasasankan@gwu.edu}
\affiliation{%
  \institution{George Washington University}
  \city{Washington}
  \state{DC}
  \country{USA}
}

\author{Christian Gütl}
\email{c.guetl@tugraz.at }
\affiliation{%
  \institution{Graz University of Technology}
  \city{Graz}
  \country{Austria}
}

\renewcommand{\shortauthors}{Rebol, et al.}

\begin{abstract} 
Augmented reality (AR) has great potential for use in healthcare applications, especially remote medical training and supervision. 
In this paper, we analyze the usage of an AR communication system to teach a medical procedure, the placement of a central venous catheter (CVC) under ultrasound guidance. We examine various AR communication and collaboration components, including gestural communication, volumetric information, annotations, augmented objects, and augmented screens. 
We compare how teaching in AR differs from teaching through videoconferencing-based communication. 
Our results include a detailed medical training steps analysis in which we compare how verbal and visual communication differs between video and AR training. We identify procedural steps in which medical experts give visual instructions utilizing AR components. We examine the change in AR usage and interaction over time and recognize patterns between users. Moreover, AR design recommendations are given based on post-training interviews. 
\end{abstract}

\ccsdesc[500]{Human-centered computing~Mixed / augmented reality}
\ccsdesc[500]{Computing methodologies~Volumetric models}
\ccsdesc[500]{Social and professional topics~Remote medicine}
\ccsdesc[500]{Human-centered computing~Systems and tools for interaction design}
\ccsdesc[500]{Human-centered computing~Interaction techniques}

\keywords{Augmented Reality, Remote Collaboration, Telehealth, Volumetric Communication}

\begin{teaserfigure}
\centering
    \includegraphics[width=0.95\textwidth]{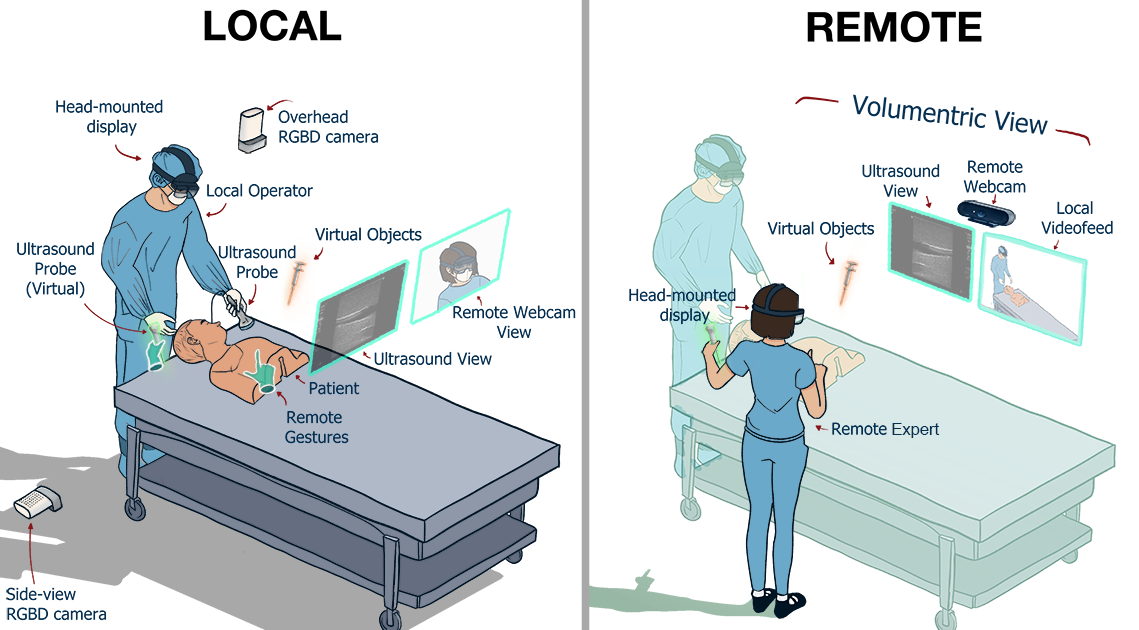}
    \caption{Augmented Reality (AR) System for Remote Collaboration. A remotely located expert (right) teaches a local trainee (left) a medical procedure, the placement of a central venous catheter (CVC). Both parties communicate in AR using head-mounted displays. The expert receives a volumetric view of the local scene and guides the trainee using gestures, augmented objects, and verbal communication. }
    \label{fig:diagram}
\end{teaserfigure}

\maketitle

\section{Introduction} 
One of the challenges in providing equitable and high-quality medical care is ensuring that practitioners are competent with relevant bedside medical procedures, which are sometimes complex and require expert support that may not always be available.
Introducing new technologies to enhance medical training and simulations, as well as their geographic reach, is one important approach to addressing this challenge. Historically, remote procedural support was provided by telephone, where an expert walks the person needing to perform a procedure through the steps using only their voice. 
Improved connectivity has led to increased availability and use of video conferencing. However, video conferencing may not be well suited in some situations that require more 3D context or "hands-on" demonstration. Augmented reality (AR) and mixed reality (MR) technologies allow for improved context of the environment and conveyance of gestural demonstrations in the performers' field of view. 
This research specifically focuses on investigating AR collaboration during a long and challenging procedural training task. 

In this work, we present the design and evaluation of an AR collaboration system 
for procedural skill training. Our AR system design choices were made in conjunction with domain experts during development. By evaluating the system, we provide answers to the research question \emph{``How can we teach procedural skill in AR?''}. We specifically focus on teaching strategies, preferred spatial workspace setups, and interaction with the user interface as well as virtual objects in AR. 

The medical procedure used to evaluate our system is the ultrasound-guided placement of a central venous catheter (US-CVC). 
Based on the challenges of the US-CVC procedure, we determine our system requirements and design in  \Cref{sec:mr-system-requirements} and \Cref{sec:mr-system-implementation}, respectively. To evaluate the system design, a research study was conducted (\Cref{sec:research-study}). Results and the discussion of human-AR-interaction principles for procedural training are presented in \Cref{sec:results-discussion}. 
Our main contributions are:
\begin{enumerate}
    \item We present AR design principles including spatial user interface and workspace setup for procedural skill training. 
    \item We demonstrate how medical experts utilize tools provided by AR to optimize teaching complex medical procedures.
    \item Develop best practice recommendations on AR teaching. 

    \item We compare visual communication and workload in AR against video-conferencing-based training. 



\end{enumerate}


\section{Related Work} \label{sec:related-work} 
We review AR that supports collaborative procedural skill acquisition in a broader sense and the healthcare domain.

\subsection{Augmented Reality Collaboration}
The first general human-computer-interaction (HCI) guidelines for AR were introduced by Dünser \etal \cite{dunser2007applying}. Although the AR software and hardware have changed significantly since then, the main concepts such as affordances, learnability, responsiveness, and feedback still play an important role in current collaborative AR systems, including our system. A more recent overview on HCI in AR is given by Janani \etal \cite{hcioverview}. 
In contrast to these generic AR HCI design principles, this work aims to provide AR collaboration design and teaching guidelines. Wang \etal \cite{WANG2021102071} provide a comprehensive overview of AR collaboration systems. They identify open issues including 3D reconstruction and multimodal interaction \cite{Kim2020} that are discussed in this paper. More recently, Marques \etal \cite{9506837} introduced a taxonomy to foster the creation of a conceptual framework for AR collaboration. Sereno \etal \cite{Sereno2022} derives design considerations for collaborative AR environments focusing on visualization. A timeline of the development of AR collaboration was presented by Bellen \etal \cite{Belen2029} and Ens \etal \cite{ENS2019}. 

Schopf \etal \cite{schopf2023affordances} identify AR affordances for co-located communication. Henderson \etal \cite{henderson2011augmented} show the superiority of AR over a traditional display when completing a procedural assembly task. Wijdenes \etal \cite{wijdenes2018leveraging} present heuristics to evaluate CVC AR software. Our work goes beyond general AR design principles and gives concrete suggestions for AR workspace setup and remote teaching for procedural training. We focus on collaborative AR system design. 

Sun \etal \cite{sun2019augmented} focus on simple gestural interaction for children's education. When creating a collaborative AR system, these gestures need to be used for visual communication between the parties. Hence, they cannot be used for user-system-interaction. We decided to rely on a hand menu activated through gesture and gaze \cite{10.1145/3530886, 10.1007/11872320_56} instead. A thorough comparison of hand interfaces was conducted by Datcu \etal \cite{doi:10.1080/10447318.2014.994193}. 

\subsection{Augmented Reality in Health Care}
Several AR systems have been developed and undergone proof of concept to teach medical procedures to trainees. 
A technical description of a real-time holographic communication system was presented by Orts-Escolano \etal in 2016 \cite{orts-escolano2016holoportation}.  
Gasquez \etal \cite{10.1145/3411764.3445576} describe telementoring procedural skills during emergent surgical procedures. The supervising surgeon is presented with a 3-dimensional reconstruction of the patient and uses gestures, annotations, and pre-recorded procedural video clips to provide instruction to the novice. There are no virtual tools with which to demonstrate the preferred relative spatial orientation of the surgical tools. Interestingly, interviews with expert surgeons suggest that having a virtual representation of an actual instrument could be beneficial in complex cases.

Yu \etal \cite{9712238} presented a co-located AR collaboration system using real-time 3D reconstruction. Strak \etal \cite{10.1145/3473856.3473883} found that their teleconsultation system improved performance during electrode placement compared to video-conferencing. Roth \etal \cite{9419320} designed their MR system for collaboration in the ICU during the Covid-19 pandemic. 
Fowers \etal \cite{fowers2022participatory} took a very similar approach building on the work of Orts-Escolano \etal \cite{orts-escolano2016holoportation}. 
These MR collaboration systems provide many new possibilities to spatially interact remotely. We identify the need to elaborate on the design choices and the collaboration methods in depth, especially during procedural training. Best practices in system design and remote spatial training will increase the success of such systems in the future.

We find the following differences when comparing our AR collaboration system to existing work. Schoeb \etal \cite{schoeb2020mixed} used a mixed reality system to teach Foley catheter placement to medical students. Their system presented students with instructional videos displayed on a virtual screen above a mannequin on which the procedure was performed. Students could select a video describing the next procedural step in the sequence. However, the system did not allow for the provision of interactive teaching from a live instructor, nor did it overlay objects or visual cues directly on the mannequin. 

Kobayashi \etal tested an MR system for CVC training that overlaid subcutaneous anatomical structures on a mannequin used to practice the procedure.\cite{kobayashi2018exploratory} While this system did allow for live instruction, it did not integrate the use of virtual objects that the instructor could manipulate and use to demonstrate procedural steps on the mannequin. Additionally, it presented subcutaneous visual information over the mannequin that would not be available to a proceduralist during a real CVC insertion in most cases. Moreover, remote US training adds additional challenges to the communication system \cite{Kessler2016_US2, article_US3}. Mahmood \etal \cite{Mahmood2018_US1} proposed how US views can be used effectively in AR.

Mather \etal conducted a feasibility study of a real-time two-way communication system for remote procedural guidance that would allow the trainee to see the instructors' hands overlaid on the real environment at the procedure site. \cite{mather2017helping} The trainee's head-mounted display (HMD) captured a 2-dimensional image of their environment that the instructor would view on a flat screen, while a camera above the instructor would capture their hand gestures for display in the trainee's HMD. 

Rojas-Muñoz \etal developed a system to allow for unidirectional interaction with a remote environment for real-time procedural guidance. \cite{munzer2019augmented} In their system, the remote instructor is presented with a 2-dimensional image of the trainee's surgical field on a flat touch-screen during a surgical procedure. The instructor could annotate the surgical field with markers and virtual objects, which are then projected in 3D over the trainee's surgical field in 3 dimensions through an HMD. This system does not present 3-dimensional information from the trainee's location to the remote instructor, and the remote instructor must therefore annotate a 3-dimensional environment on a 2-dimensional plane. 

A similar AR collaboration system for first aid assistance has been designed and evaluated in a simulated emergency by Rebol \etal \cite{rebolCPR}.

Overall, we find that existing AR collaboration research does not investigate long, complex, stressful procedural tasks and how AR artifacts are used for teaching as well as perceived by the learner. 
Moreover, existing AR systems with volumetric capture are not portable enough to support a variety of procedures from different domains.

\section{AR System Requirements} \label{sec:mr-system-requirements}
We determined that the US-CVC procedure, the placement of a vascular access catheter into large veins that empty directly into the heart, has certain features that make it a good candidate for this research. US-CVC is a complex multi-step procedure that requires bimanual dexterity, visual attention away from the hands, and 3-dimensional tool alignment. It is a critical but low-frequency skill. US-CVC also requires a higher resolution than many AR-guided procedures reported in the literature.

We started the requirements analysis with an elicitation study in which we recorded ten in-person US-CVC training sessions on a mannequin in a simulation setting, each of which lasted between 50 and 100 minutes. Detailed elicitation study results including a first evaluation phase are presented in \cite{rebol2023collaborative}. The analysis helped us to better understand the medical procedure and to specify the features of an AR training system. We built a prototype based on the initial requirements. For two years, we periodically asked four US-CVC experts for feedback and added and removed features to improve the system design. In between the discussions, the US-CVC experts were able to test the system in short sample training sessions. On the highest level, we identified the need for both audio and visual communication. 
We group the requirements for visual and non-verbal communication into three categories:

\paragraph{Volumetric data}
The remote expert needs to get volumetric data from the local room to understand the exact 3D positions of the US probe and needle relative to the patient to facilitate optimal skin entry position and angle. Similarly, the local trainee may benefit from receiving instructions using volumetric information, to enhance communication of spatial orientation and movement of the probe and needle in an unconstrained system. This is consistent with best practices for the demonstration of open, multidimensional complex skills \cite{https://doi.org/10.7863/ultra.33.8.1349}, and may help to reduce the cognitive workload of the trainee by decreasing the need to interpret verbal instruction \cite{doi:10.3109/0142159X.2016.1150984}. 

\paragraph{Objects and annotations}
Virtual objects replace physical objects in AR. The remote expert can demonstrate spacial concepts as well as show where and how to use tools, similar to in-person training. Procedure-relevant objects need to be modeled or scanned before the AR system is deployed. Drawing and pointer annotations may be useful, although there is no similar method of communication that could be identified during in-person training.

\paragraph{Gestural communication}
Gestures are needed for effective visual communication. Both communication parties need to be able to point at landmarks. Moreover, other types of gestures such as optimal hand position \cite{types-gestures-1992} need to be conveyed during medical training. Besides communication, hands, and arms are important when demonstrating the usage of tools.  


\section{AR System Implementation} \label{sec:mr-system-implementation}
We designed a two-party augmented reality communication system for medical training. A technical description including implementation details had been published in \cite{rebolISMAR}. The incremental design decisions were made together with domain experts.
Four US-CVC experts and study instructors had been testing the system during development and provided valuable feedback in the design discussions. After we designed the basic system, the domain experts evaluated their preferred workspace setup (\Cref{sec:workspace-analysis}) in short sample training sessions. 

\subsection{Design Decisions}
The major design decisions related to the volumetric view capture, the interaction between the users, and the interaction of the user with the AR system. The volumetric view capture was the most challenging aspect due to the need to satisfy the requirement of having a lightweight system and setup while providing enough precision and visual information for medical procedures in which details and accuracy are crucial. 

\paragraph{Volumetric view and cameras} During the design discussions, we found that the accuracy from state-of-the-art consumer RGBD cameras \cite{rgbd-comp, rgb-comp2} is not sufficient for very small and translucent objects during parts of medical procedures. Thus, we use large 1920x1080 screens in addition to 3D scene reconstruction in our system. The 3D mesh facilitates spatial understanding and aligned interaction. Whenever the 3D mesh is lacking detailed information because of depth sensor limitations, the user can fall back to augmented two-dimensional screens to get the information needed in high resolution. 


Moreover, real-time point-cloud matching is not fast and precise enough to provide a high-quality volumetric view using two cameras. We require a lightweight setup with only two cameras to allow for flexibility between procedures and fast deployment of the system. Hence, we alternate between 3D views. Whichever camera angle is closer to the angle at which the user observes the scene is rendered. Our tests showed that automatic switching between cameras can be disorienting for the user. We allow the user to manually switch between the camera positions instead. Manual switching is also necessary because people and objects may move and block camera angles during different parts of the medical procedure. 


\paragraph{User-to-user interaction} We found that augmented tools in combination with augmented hand models are most efficient for teaching the CVC procedure. We experimented with surface drawing and pointing tools, which have been used in many AR schemas. Surface drawing allows the remote expert to draw on the 3D mesh, and surface pointing highlights an area that is visible when the expert’s index finger moves close to the mesh. However, many of the tasks demonstrated by experts during US-CVC required spatial positioning away from, or at angles to the mesh surface - both difficult to render with surface drawing tools. In addition, teaching tool handling and movement required complex hand movements that triggered pointing when not desired, which was distracting to trainees. Instead, the use of virtual 3D task-specific tools to demonstrate orientation relative to the mesh surface, and accurate representation of hands and hand movements were preferred by both experts and trainees. This finding suggests that mapping typical expert gestures during teaching, and ensuring that fidelity of critical motions is preserved in the teaching space, are essential in developing the expert's augmented teaching toolkit.

\begin{figure}
    \centering
    \includegraphics[width=\columnwidth]{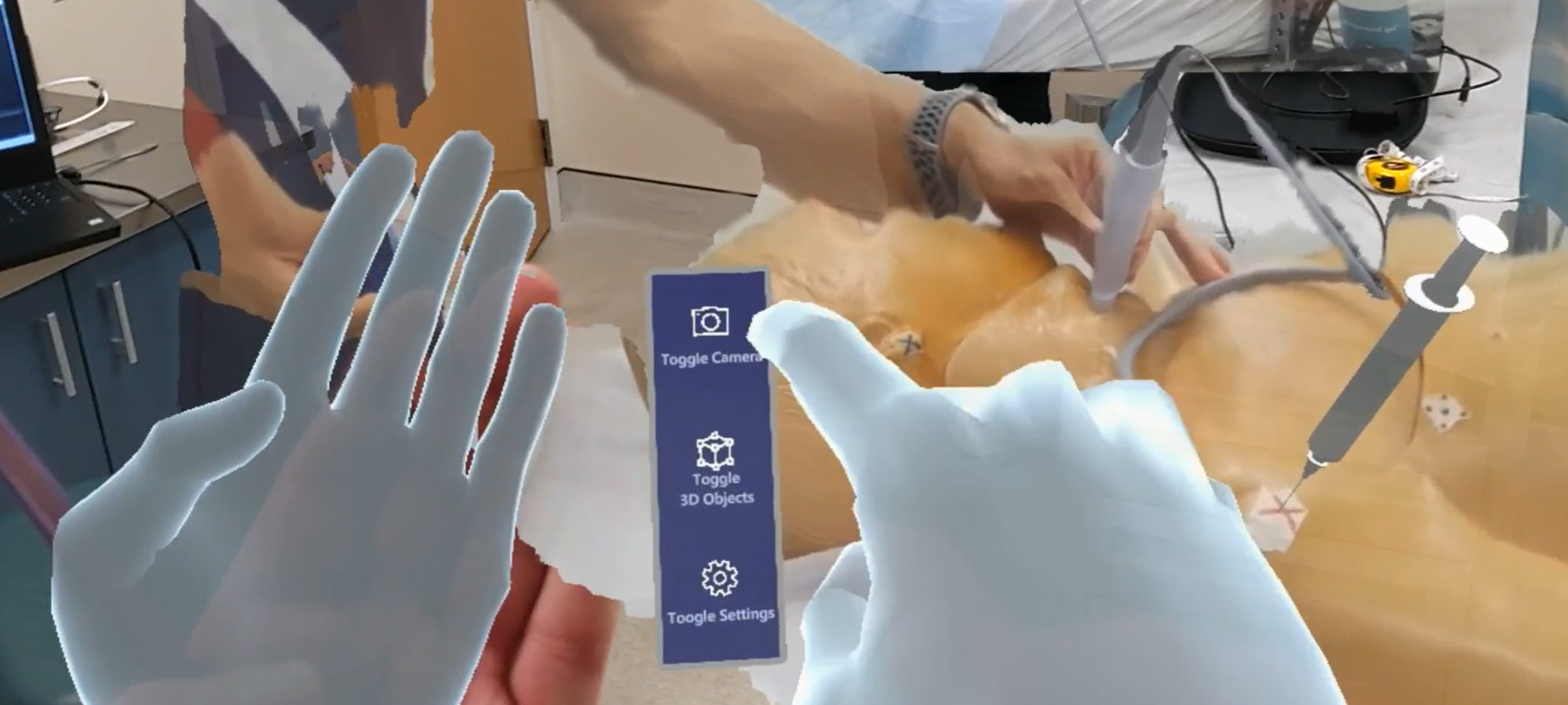}
    \caption{Hand Menu. The user can switch between cameras, toggle objects, and expand settings.}
    \label{fig:hand-menu}
\end{figure}

\paragraph{User Interface} Periodically, users are required to interact with the AR headset during training. Ideally, this should be a seamless device-user interaction that minimizes distractions, allowing the user to focus on the primary task. The four US-CVC experts tested four methods of user interaction with the AR headset during the design phase - voice, gesture, anchored menus, and hand menus. The tasks included system setup by placing virtual balls on top of physical markers, connecting to users, switching camera views, and toggling gesture representations. The evaluation method included observation and interviews. 
We use the anchored menu during system setup which consists of connecting the devices and aligning the views. The anchored menu consists of button groups needed for setup and can be moved to the desired position. The anchored menu collapses after the initial setup because it is not needed during medical procedure training. The hand menu opens when the user looks at their palm as shown in \Cref{fig:hand-menu}, and allows the remote expert to toggle camera views and visibility of augmented tools during the procedure. After a short learning curve, experts were able to easily toggle between camera views and engage virtual tools as needed. 

Experts found that gestures and voice commands to control user interfaces interfered with teaching activities. The "air tap" gesture, 
a pinch gesture in front of the headset was initially selected to switch between camera views. However, the air tap was similar to hand movements used during task demonstration, and would inadvertently switch cameras during periods of gestural teaching. Double air taps, although sufficiently different from teaching gestures that they did not occur inadvertently, required multiple attempts to be recognized, causing experts to miss seeing critical actions due to delayed camera switching. The use of voice commands resulted in similar problems. The voice commands were triggered by accident during the training as the experts and trainees actively communicated verbally throughout the procedure. Furthermore, trainees were distracted by expert voice commands not directed at them. These findings suggest that actions that are used only by the expert for local environment control should be triggered in a way that is not communicated to the trainee.

\begin{figure*}
    \centering
    \def\figoneheight{2.66}
    \def\figtwoheight{2.47}
    \begin{subfigure}[b]{\textwidth}
         \centering
         \includegraphics[height=\figoneheight cm]{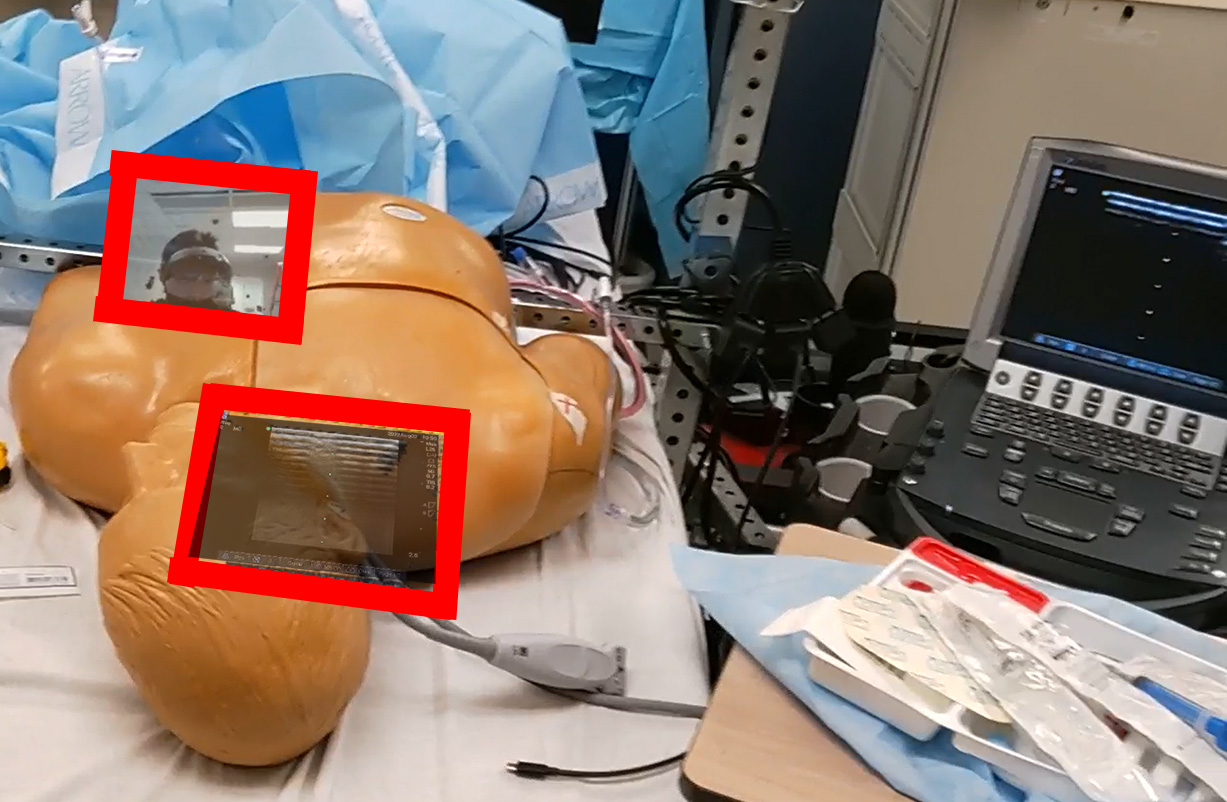}
    \includegraphics[height=\figoneheight cm]{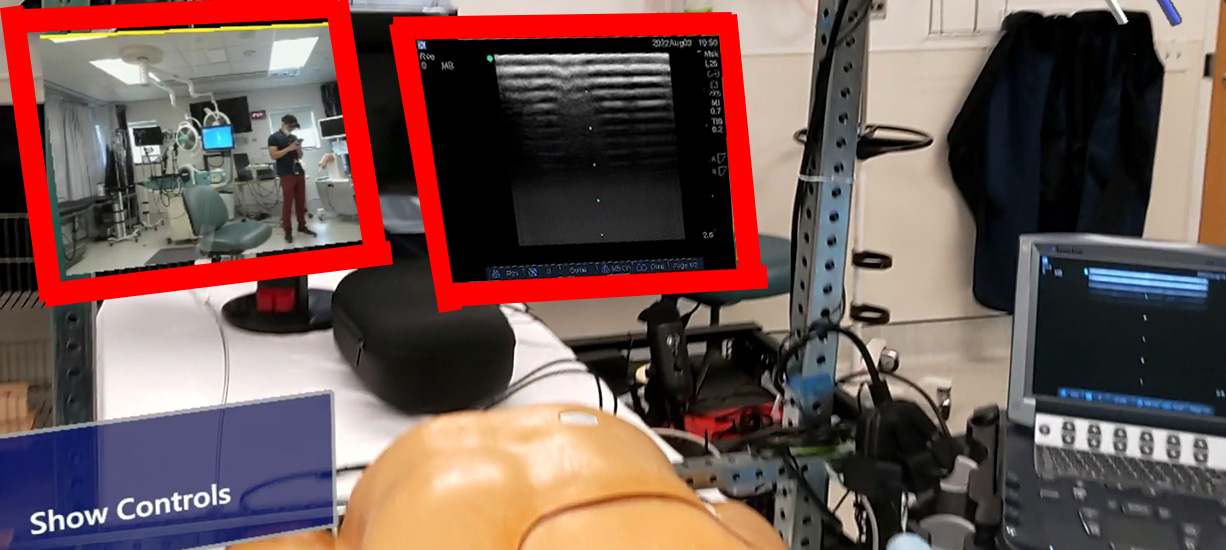}
    \includegraphics[height=\figoneheight cm]{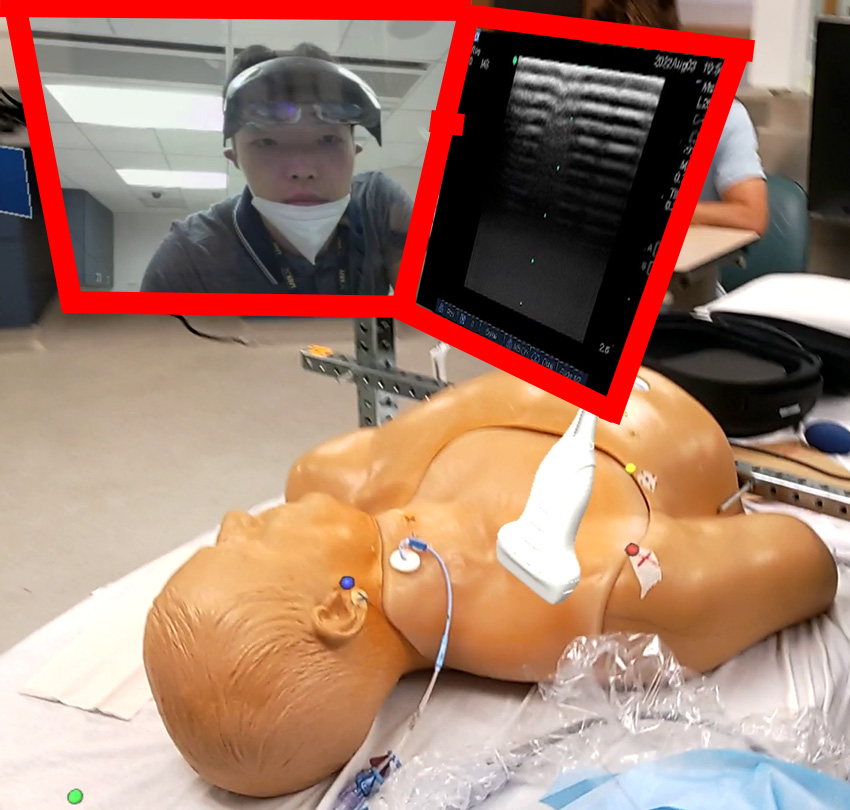}
    \includegraphics[height=\figoneheight cm]{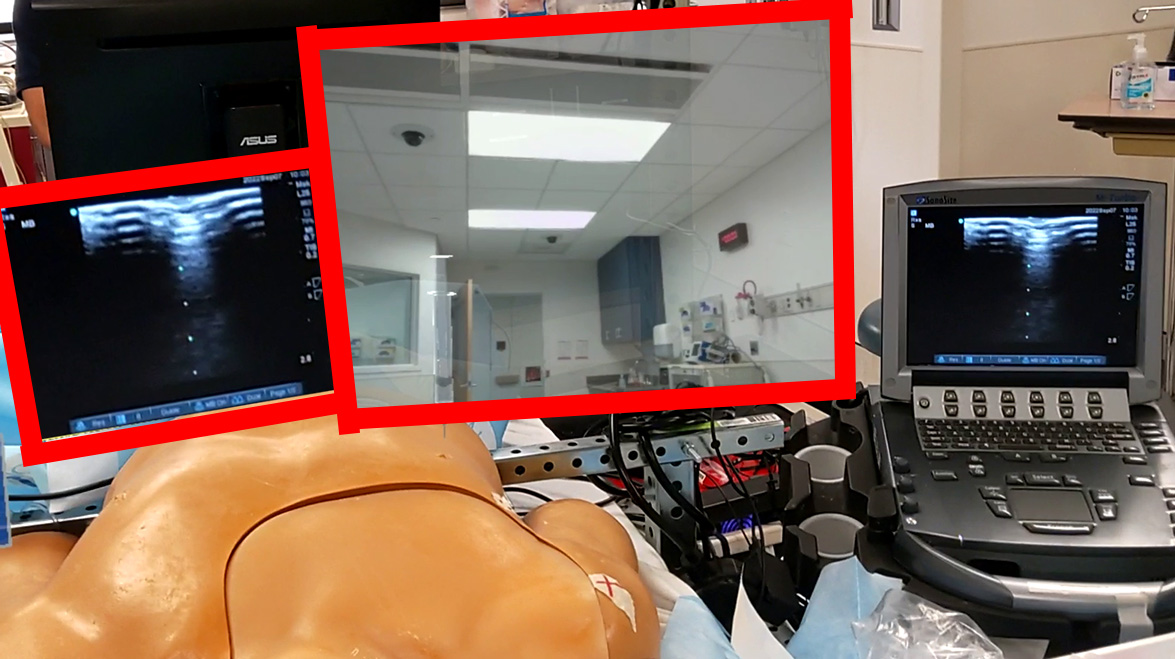}
    
         \caption{Trainee workspace}
         \label{fig:workspace-trainee}
         \vspace{0.3cm}
     \end{subfigure}
     
    \begin{subfigure}[b]{\textwidth}
         \centering
    \includegraphics[height=\figtwoheight cm]{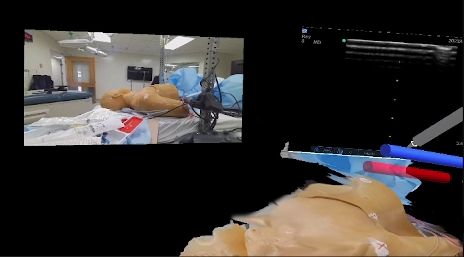}
    \includegraphics[height=\figtwoheight cm]{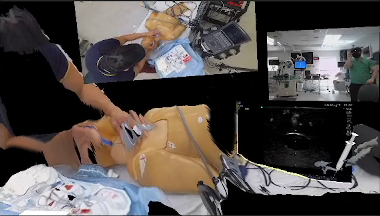}
    \includegraphics[height=\figtwoheight cm]{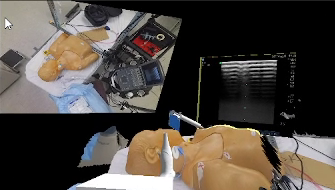}
    \includegraphics[height=\figtwoheight cm]{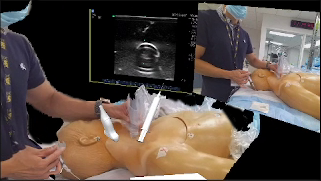}
             \caption{Expert augmented workspace}
         \label{fig:workspace-instructor}
    \end{subfigure}
    \caption{Augmented workspace for US-CVC. Four US-CVC experts present their preferred trainee (a) and expert (b) workspace setup. We highlight the augmented US and webcam screens in the trainee workspace using red borders. The expert workspace consists of the volumetric mesh, 2D screens, and virtual objects. We only show the augmented workspace without the environment for presentation proposes.    }
    \label{fig:augmented-workspace}
\end{figure*}

\subsection{Augmented Views}
The system was designed asymmetrically such that the trainee's focus is on the physical workspace, with augmentations used to enhance training. In contrast, the expert's view is dominated by augmented information, with physical world interactions limited to a procedure checklist and their own physical space. We illustrate the views in \Cref{fig:augmented-workspace}.
 
The communication roles are asymmetrical. The instructor or remote expert sees a virtual view of the local scene which includes the local trainee, the patient, the tools needed for the procedure, and the US machine. The scene is presented to the remote expert using a volumetric view of the procedural field, combined with an augmented 2D video screen of the larger room, and a live US video feed. In addition to observing the procedure, the remote expert can guide the local trainee through the movement of virtual 3D objects and virtual hand gestures that are augmented into both the expert and trainee's fields of view. 

The trainee's view is dominated by the physical environment, which includes the patient, toolkit, and US machine. The local trainee receives augmented input in the form of a 2D video screen of the remote expert, and a virtual representation of the live US feed that can be placed anywhere in the field of view. The local trainee receives augmented gestural and virtual object guidance controlled by the remote expert. 
Both trainees and experts can place virtual objects at desired locations within their workspace.

\subsection{Devices and Software}
The two users of the AR system connect using one Microsoft Hololens 2 (HL2) HMD each. Each user station requires a computation computer with state-of-the-art consumer hardware at their site. 
We use two Microsoft Azure Kinect RGBD cameras to capture a volumetric view. On the software side, Microsoft Mixed Reality Toolkit for Unity was used to implement the user interface. For off-device rendering that results in less computational utilization of the HL2, the Holographic Remoting Player was used. Peer-to-peer, as well as sever-client-based Websockets, were implemented to handle network communication between the devices.

\subsection{US-CVC System Setup}
We illustrate the system setup for US-CVC in \Cref{fig:diagram}. 


After initial training attempts in which remote experts were using their hands to simulate US probe and syringe/needle complexes, US-CVC-relevant 3D objects were added to allow the remote expert to demonstrate concepts and the usage of medical tools. Simple shapes were adequate to represent some objects, \eg the vein and the artery. However, more complex shapes, for example, a US probe with a directional tab, and a syringe/plunger tool used to demonstrate proper handling required more detailed virtual tool development.


\section{Research Study} \label{sec:research-study} 
We conducted an Institutional Review Board-approved research study to evaluate AR teaching of procedural skill. We focused on the US-CVC procedure as an example of a high dexterity, complex, multi-step procedure that stresses the limits of the AR headset and 3D capture technology. We designed the study such that the AR teaching takes about one hour. The multiple-step simulation gives insights into how AR can be used in procedural training.

\subsection{Study Design}
We are investigating one-on-one teaching of procedural skill in AR. We identify common patterns and how different aspects of AR are used. We compare teaching in AR against videoconferencing-based teaching and identify the strengths and weaknesses of AR. Although the AR system is designed for remote assistance, we run the US-CVC study locally in a medical simulation center with the student and teacher in separate rooms.

\subsection{Instruments}
We videotaped study sessions and analyzed how AR was used by watching and abstracting the recordings. In addition, experts and trainees completed NASA TLX \cite{hart2006nasa} as well as SIM TLX \cite{Harris2020} surveys, and were interviewed after the sessions. 

\subsection{Procedure}
Prior to the study, trainees and experts provided informed consent. The trainees completed US-CVC pre-training to familiarize themselves with the steps of the procedure. 
The participants completed a pre-training survey which included demographic and prior experience questions. At the beginning of each training session, experts oriented the trainees regarding the US-CVC procedure and provided background information. Then, they prepared the trainee's workspace for the procedure and talked through the medical equipment necessary for the procedure. After the initial preparation, experts moved to a separate room to start with the video or AR training. 
In the case of AR sessions, both experts and trainees completed a 5-minute AR briefing. Apart from the briefing, the participants did not receive any training on the technology. Then, the actual training session, which took about one hour, started. After the US-CVC session, both trainees and experts independently completed surveys and interviews. 

We prepared a Blue Phantom ultrasound central venous access training mannequin \cite{mannequin}, a CVC kit, and a Sonosite M-Turbo Ultrasound system \cite{us}. 
  The following parts of the CVC procedure were taught over AR and video:
  \begin{enumerate}
      \item A talk through the procedural steps, the preparation of the CVC kit, and the use of the US.
      \item US-guided venous access with a large bore needle, guidewire insertion through the needle, US confirmation of wire location, and catheter placement over the guidewire (Seldinger technique). 
      \item Flushing and drawing of three catheter ports after insertion.
  \end{enumerate}
The US-CVC procedure is performed at the bedside to gain central venous access for a variety of clinical indications.

\subsection{Participants}


We selected 25 medical trainees (7 male/18 female, on average: 27 years old, 2 years clinical experience, no AR-VR experience) and 6 medical experts (5 male/1 female, 1+ years of CVC teaching experience, average 41 years old, none to minimal AR-VR experience). 
The trainees and experts were randomly assigned into 15 AR teaching and 10 video teaching pairs.

\section{Results and Discussion} \label{sec:results-discussion} 

Results to our initial question, \emph{``How can we teach procedural skill in AR?''}, can be grouped by workspace design, system interaction, and teaching practices. Moreover, the AR teaching style was compared to video conferencing. 

\subsection{AR Workspace Analysis} \label{sec:workspace-analysis}
Outside of our regular study sessions, we asked four US-CVC experts and study instructors for their preferred augmented workspace setup for both expert and trainee workspaces. We show the results for the trainee and the expert view in \Cref{fig:augmented-workspace}. 

\paragraph{Trainee workspace.} Three US-CVC experts positioned the two screens above the physical workspace. One expert positioned both displays very close to the area of operation to ensure that the screens were in the field of view (FOV) while operating. The screen sizes in this setting are smaller. The ideal AR training workspace has to provide a clear view of the operating field. It is beneficial to have procedure-relevant information and screens close to the operating space such that the trainee has additional information in their FOV. Current FOV limitations of the HL2 constrains placement to a small area around the active procedure. Recently introduced HMDs offer a larger FOV. 

\paragraph{Expert workspace.} All four experts positioned the volumetric view horizontally below the screens. The screen size between configurations is similar. Two experts rotated the screens such that they can teach from the right side of the mannequin, similar to in-person training. The other two experts rotated the screens such that they could teach from where the trainee stood. 
Although the majority of the experts decided that they would want to teach the procedure from a similar position to where they would stand when teaching in-person, we find that this is not the best choice as shown in \Cref{sec:ar-teaching-practicies}. Trainees prefer to receive instructions from their point of view (POV). An AR tutoring system has to be flexible regarding the workspace setup to allow for the best learning experience. 

\subsection{AR System Interaction}
After each US-CVC training session in AR, we interviewed both the trainee and the instructor to facilitate a holistic understanding of the AR training experience. In addition to the interviews, video recordings were analyzed to identify patterns between sessions.  




\paragraph{UI Menu Interaction}
The experts and trainees used the hand menu and anchored menu, respectively, as described in \Cref{sec:mr-system-implementation}. 
%
Although multi-modal interaction can be beneficial \cite{WANG2021102071, Kim2020} it has to be implemented taking the application domain into account.
Distractions were caused by the built-in UI configurations. Accidentally triggered voice commands caused the HL2 to show debugging information during the procedure, which needed to be deactivated using additional voice commands. The HL2 start menu, activated by a wrist-tap gesture, was triggered when the local trainee donned sterile gloves. At other times the eye-gaze open triggered menu activation. Each of these events was visually distracting to the users, necessitating a pause in training to correct it. 
While teaching users how to recover from accidental HCI interactions is useful, a better practice may be to deactivate voice and diagnostic capabilities before deploying AR software. 

\paragraph{Hand Tracking}
Although experts reported that they could demonstrate larger hand motions confidently, the HL2 hand tracking did not provide precise enough hand tracking to capture all the hand poses the experts wanted to demonstrate. US-CVC requires precise individual finger movements for select tasks, such as manipulating a syringe or holding the US probe. Because the virtual hand is rendered from the remote expert's HL2 camera, parts of the hand not directly visualized are not adequately tracked. To alert the remote expert to occlusions, the virtual hand model was augmented onto the expert's FOV. In this manner, the expert was able to see the hand position presented to the local trainee and adjust positioning to optimize the demonstration.  A dedicated hand-tracking system would improve the virtual representation but at a computational cost. 
Thus, it is important to match the desired hand-tracking precision to the specific training need. Furthermore, allowing the expert to visualize not only their hand, but the representation presented to the remote trainee can enhance gestural communication even with limited tracking precision.

\paragraph{Object Manipulation}
We discovered a learning curve for the experts manipulating augmented objects. During the first session, the moving and resizing required several attempts, which improved with practice. The learning curve was similar for screen adjustments, \ie positioning and resizing. 
We noticed that experts were not aware of the limited area in which the hand is tracked. Sometimes the hand left the tracking space while an object was manipulated. Hence, the hand tracking stopped and the object remained at an undesired location. We recommend reminding AR users in the pre-session briefing that they need to look at the object while manipulating it to make sure the hand-tracking sensors capture the hand. This is HL2 specific because it only has top and side-facing tracking cameras, compared to other headsets that have cameras that point downwards to capture the user's gestures in a larger region.
The default MRTK object manipulation suggests the use of one hand to move and rotate an object and the use of two hands to resize an object. Some participants grabbed objects with both hands when trying to move and rotate them. Thus, whenever scaling is not needed, two-handed moving and rotating should be enabled.

\paragraph{Visual Focus}
Experts alternated between the volumetric view, the video screen, and the US screen depending on the situation. During needle insertion and wire confirmation, the experts mainly looked at the US screen. During the rest of the procedure, the experts utilized the volumetric view, especially to visually guide the trainee, and the video view to get a detailed view of the instruments used. While specific positioning can be left to the individual, the AR meshes, objects, and screens should always be close together such that changes in focus can be made quickly. 

The trainees' gaze was directed primarily at the patient at the site of catheter insertion, 
 with intermittent brief focus shifts to the expert webcam screen. During needle insertion, the trainees could view either the augmented US screen or the actual US screen. About half of the trainees used the augmented US screen because they could place it closer to the area of operation, 
while the remainder used the US machine display because there was no noticeable delay between hand movements and corresponding changes in the US image. We find that it is sometimes important to provide alternatives such that the user can decide if they want to use an augmented or a physical element depending on their personal preference. 

The experts noticed a better volumetric mesh quality before the draping of the patient. The blue surgical drape provided less contrast than the orange mannequin which made it more difficult to distinguish the instruments on top of the drape. Moreover, the uneven surface in combination with the transparent surfaces of the drape results in low volumetric mesh quality. We learned that the surface property of items needs to be taken into account when used for time-of-flight RGBD camera capture.


\begin{figure*}
    \centering
    \includegraphics[width=1.0\textwidth]{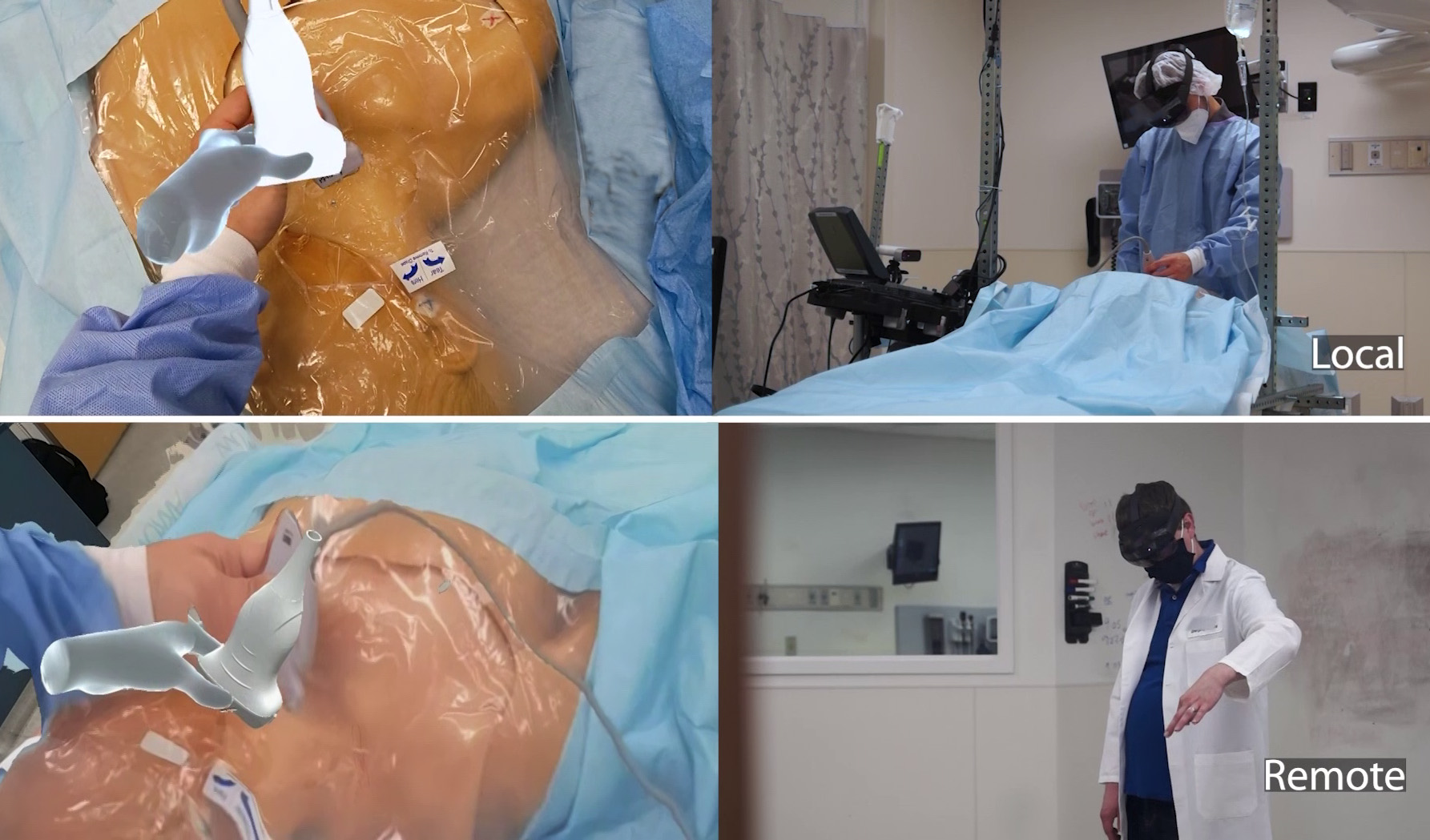}
    \caption{Teaching in AR. The remote expert (below) uses his hands and virtual tools to guide the local trainee (top) directly on top of the US probe. In AR, the expert can stand in the same position as the trainee. Note that augmented visuals appear more transparent through the head-mounted display than in the images.}
    \label{fig:teaching-on-top}
\end{figure*}

\subsection{AR Teaching Practices} \label{sec:ar-teaching-practicies}
We derive best practices for AR teaching from post-training interviews and observations during the training. We found that many of these practices are not only relevant for specific procedures but for teaching and learning in AR more broadly. 

\paragraph{Positioning When Co-Located in AR}
We found that during most of the training, experts position themselves to the trainee's right at a 90-degree angle, the location typically used for in-person training. In the physical world, this location provides a clear view of the angle of needle insertion during the US-CVC procedure and access to procedural tools. 
Some experts also taught from the head of the bed, where the trainee was standing, or directly behind the trainee's shoulder. This enabled the expert to have the same point-of-view (POV) as the trainee without interfering in space and allowed the expert to be closer to the task. Experts found that the resulting view from the position of the operator was likewise intuitive. Furthermore, it aided experts in understanding how the learner was manipulating tools. Interestingly, most experts did not immediately recognize the benefit of co-located POV training and noted a learning curve on optimal teaching positions. 
The trainees prefer receiving instructions from their POV, where expert hand, arm, and tool position can most closely demonstrate the desired skills. We recommend pretraining guidance for remote experts to consider local operator perspectives when demonstrating skills.

\paragraph{Spatial Teaching}
Experts and trainees reported that the concept of teaching in AR directly on top of the physical instruments without interfering with the trainees' space could make teaching more effective. For example, the experts used the augmented US probe object to gradually correct the position of the trainee's probe as shown in \Cref{fig:teaching-on-top}. The AR tools were also used to explain the relative position of anatomy in space which allows trainees to understand how to utilize the safest approach to the procedure. This was distinct from the way an expert might correct a trainee in co-located training because the expert may demonstrate the proper movements in the actual field of work without direct inputs to the US probe. The experts could give spatial anatomy instructions, such as visualizing the blood vessels inside the simulation mannequin using 3D models. This new concept of spatial teaching has the potential to facilitate better understanding and cannot be achieved when co-located due to physical constraints. 

\begin{figure}[]
    \centering
    \includegraphics[width=\columnwidth]{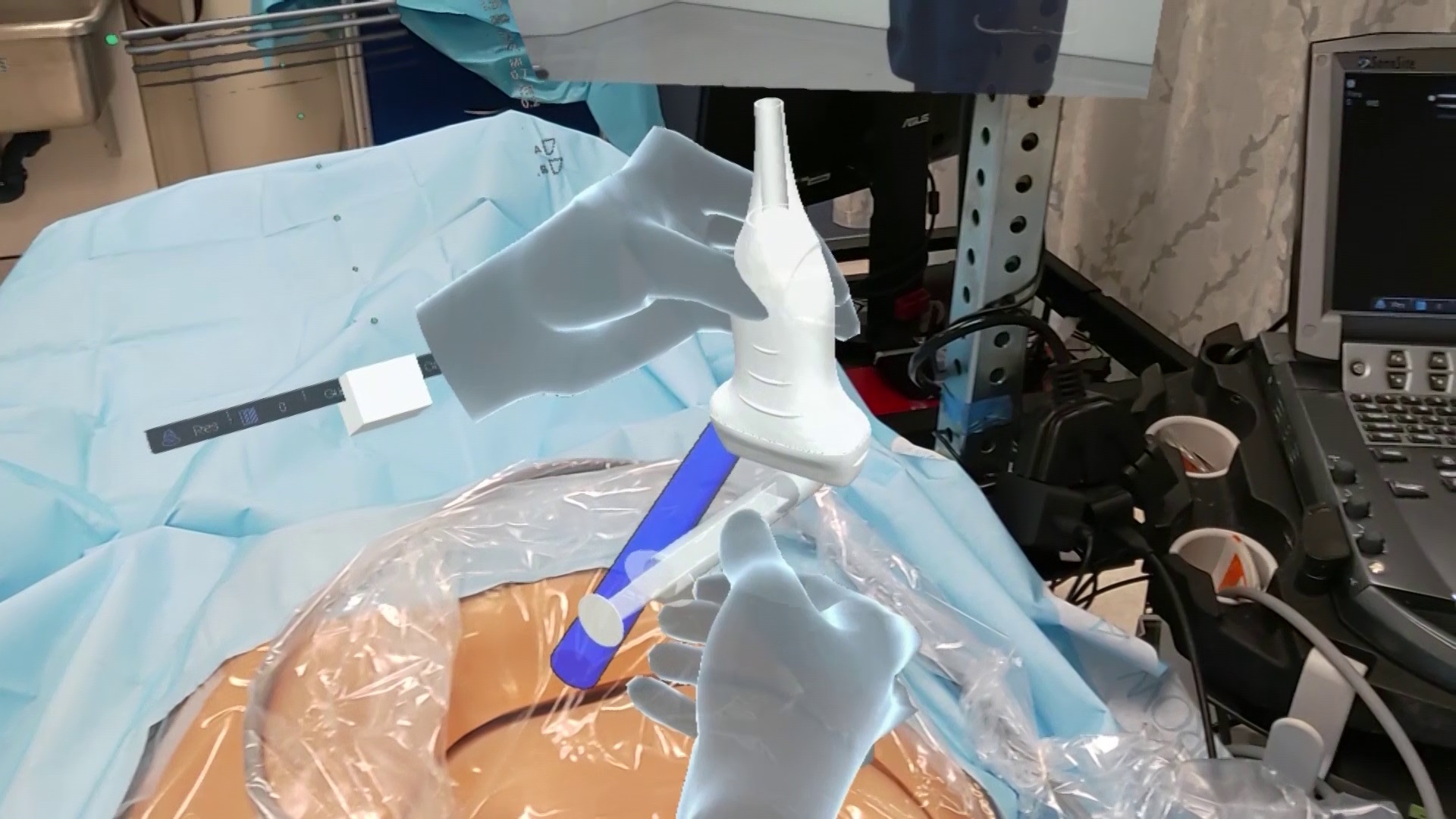}
    \caption{Trainee View. The remote expert uses her hands and virtual tools to explain a concept in the trainee's field of work.}
    \label{fig:teaching-fow}
\end{figure}

Trainees reported that they prefer to receive augmented spatial demonstrations from experts close to their field of work as shown in \Cref{fig:teaching-fow}. It allows them to remain focused on their task while receiving additional input. Trainees had to be reminded that visual demonstrations were given to them when they were far away from their field of work because of the limited FOV of the HL2.

\paragraph{Types of Visual Instructions}
The three most dominant and effective visual teaching methods in AR consisted of:
\begin{itemize}
    \item Gestures with hand tracking of the remote expert and a hand model representation for the local trainee.
    \item 3D models moved over real tools to give spatial directions.
    \item 3D models in combination with hand models to demonstrate procedural concepts and the usage of medical tools. 
\end{itemize}
The remote expert used pointing gestures on the augmented screens and on locations in the local trainee's environment which are presented as a volumetric model on the remote side. The pointing gestures helped especially during times when verbal instructions were unclear.
Trainees reported that the augmented hand model and tools could be distracting when not in use. 
In addition, hand position tracking was intermittent when hands were not in the HL2 FOV, resulting in augmented hands that stopped moving and then appeared elsewhere in the learner's FOV. Local trainees were often not sure which hand movements were relevant. 
We recommend that the expert remove hands from the HL2 FOV when not in use, and that hand tracking is enabled only during active teaching.

The experts utilized virtual 3D objects to show trainees how to hold, place, and manipulate tools, and to teach relevant anatomic material. Concepts such as the patient's arterial and venous anatomy were explained using abstract 3D models, which experts positioned above or next to the active workspace. Tool models with realistic essential features were used to demonstrate hand positioning and movement. For example, experts used a syringe model with a plunger to demonstrate different handholds commonly used during the procedure. Tools were also used to demonstrate the correct positioning and movement on the patient. During the US evaluation of the patient, the trainee could mimic the virtual US probe movement on the neck and observe changes in the local US image. The trainee could follow the trajectory of the augmented 3D model with the actual tool such that it was positioned correctly. 

Virtual demonstrations of anatomy, tool handling, and movement were best visualized when near the procedural field, as they remained in the trainee's FOV. Some trainees did not see virtual demonstrations that were located away from the patient, and experts had to explicitly redirect attention to the tool location. In other cases, experts left tools and hand models in the trainee's FOV overlying the patient after the demonstration, and trainees had to ask them to remove the items for better task visualization. Given the limited FOV of HL2, we recommend that experts use virtual tools on or directly adjacent to the trainee's procedural workspace, and explicitly identify when demonstrating skills away from the procedural field. In addition, experts should remove their hands and tools from the workspace when guidance concludes to avoid cluttering the learner's FOV.

\paragraph{Learning in AR}
Overall, trainees reported that they felt less pressure learning when the expert was not co-located. This separation allowed them to better focus on the steps and the procedure without fearing that the expert would assume control and 'take over" the procedure. Trainees generally felt that the expert was able able to assist if needed via the augmented system. In rare cases where trainees found tool movement or coordination concepts difficult to grasp due to previously described limitations, they indicated that a co-located expert would have been helpful.

\paragraph{Most Important AR Teaching Features}
When asked what additional features experts would welcome, most mentioned more procedure-specific and dynamic (\eg syringe with working plunger) virtual objects. Experts requested a feature that would allow them to reset the position of the augmented objects 
to avoid having to search the workspace. We recommend that the AR system provide a mix of tools that includes abstract models for flexible expert-driven demonstrations, as well as realistic, procedure-specific tools tailored to essential task performance.  

\subsection{AR vs Video-conferencing}
The results of the NASA and SIM TLX instruments are presented in \Cref{tab:workload}. No significant difference in workload was found when comparing the AR with the video group. We conclude that the workload is dominated by the procedure rather than the technology. 

Moreover, we evaluated the extent of verbal and visual communication during US-CVC procedure training. Although some tasks, such as US needle tracking, were highly gestural in both domains, we identified three procedural steps in the AR setting where experts used visual communication, \ie AR tools, more frequently compared to the video-based training. 
These steps were 
 \begin{enumerate}[(a)]
     \item How to hold a US probe, 
     \item Initially showing how to hold and draw back on the syringe, 
     \item Dilating down to the vessel using a gentle twisting motion, holding the dilator near the skin. 
 \end{enumerate}
This suggests that AR promotes the use of gestures and visual teaching beyond that which is done via video, which may enhance learning by providing another modality for information transfer. In addition, the availability of augmented tools can motivate experts to utilize visual teaching in the AR space. 






\begin{table}[]
	\setlength\extrarowheight{2pt} 
	\setlength{\tabcolsep}{10pt}
	\centering
 	\caption{NASA TLX and SIM TLX workload for trainees and experts during US-CVC training using AR compared to videoconferencing technology.
	}\label{tab:workload}
	\begin{tabular}{r||c|c||c|c}
		&\multicolumn{2}{c||}{\textbf{NASA TLX}} & \multicolumn{2}{c}{\textbf{SIM TLX}}  \\ 
		$[\mu, \sigma]$&Video & AR & Video & AR\\ \hline \hline
		\textbf{Trainee} & $64, 10$ & $65, 8$ & $49, 18$ & $52, 18$\\ \hline
		\textbf{Expert} & $63, 13$ & $66, 13$ & $48, 18$ & $55, 15$\\ \hline \hline
	\end{tabular}

\end{table}

\subsection{Broader Impact}

We argue that the lessons learned from our study can be transferred to other types of procedural skill training and real-time assistance tasks. Both training a person remotely and assisting them in real-time require communication of the same core information and manipulations, but the interaction patterns may be different. For example, POV, positioning, and the mechanism for expert intervention are different in AR. Allowing for the expert to have the aligned POV of the trainee may have significant advantages for remote guidance and skill acquisition over even in-person teaching where this POV is not possible. Overall, virtual procedure-specific objects in combination with hand models that are spatially aligned with the learner's real-world environment were the most important visual and non-verbal tools when teaching a long, complex, multi-step procedure.

AR system interaction could be used outside of medicine for teaching and procedural guidance for tasks that require interplay between cognitive processing and manual hand movements, such as machinery repair. A remote expert could, for example, assist a local trainee with the maintenance of immovable equipment in austere environments, or with repairs of large mobile equipment such as airplanes or ships.

Our research has identified that the manipulation of views from both the instructors' and trainees' perspectives can be optimized by the individual to meet their preferences. The ability to in effect create a heads-up display for the US screen more directly into the field of view of the expert or trainee can be very impactful and potentially reduce extraneous motion and cognitive load. Teachers may require special training to optimize communication for AR teaching. Trainees may also require an orientation to take full advantage of the learning features of AR may afford.

\section{Conclusion and Future Work}
We evaluated an Augmented Reality (AR) communication system for medical training. The focus was on medical training of the US-CVC procedure. Based on an elicitation study on in-person US-CVC training, we designed high-level feature requirements for the AR system. The feature requirements were iterated upon and implemented in consultation with medical experts. Participants evaluated the implemented system in a broad research study. We videotaped the study for later in-depth analysis. The results of the analysis showed how visual communication is used for procedural training specifically in AR. We discussed the usage of and identified best practices for teaching procedures through AR. 
We identified preferred visual teaching locations and how methods of instruction should be given in a collaborative AR environment. 

In future work, we plan to evaluate the effect of teaching through AR on the trainee's skill acquisition by assessing long-term skill retention.
We will measure cognitive load more objectively using physiological signals \cite{weronika1, weronika2} to determine patterns in cognitive overload and use that information to optimize teaching and the presentation of information. This can lead to more effective learning through AR.
Moreover, we will investigate how avatar representations and synthesized gestures \cite{gesturesRealTime, gesturesFull} affect the learning process as suggested by \cite{WANG2021102071}. We will examine AR teaching in non-medical training to verify the observations. 

\begin{acks}
The work is supported by National Science Foundation grant no. 2026505 and 2026568. 
The authors wish to thank Bo Lee, Anna Hu, Coroline Bereuter, Meagan Couture, Becky Lake, Scott Schechtman, and Rahil Ashraf for their help.
Moreover, the authors would like to thank medical students and residents for participating in the research study.
\end{acks}

\bibliographystyle{ACM-Reference-Format}
\bibliography{paper}

\end{document}